\documentclass[twocolumn,showpacs,preprintnumbers,superscriptaddress,amsmath,amssymb]{revtex4}
\usepackage{graphicx,amssymb,mathrsfs,bm}

\newcommand{\be}{\begin{equation}}
\newcommand{\ee}{\end{equation}}
\newcommand{\bea}{\begin{eqnarray}}
\newcommand{\eea}{\end{eqnarray}}
\newcommand{\bsube}{\begin{subequations}}
\newcommand{\esube}{\end{subequations}}

\newcommand{\Eq}[1]{Eq.\,(\ref{#1})}

\newcommand{\la}{\langle}
\newcommand{\ra}{\rangle}
\newcommand{\ti}{\tilde}

\newcommand{\rmL}{{\rm L}}
\newcommand{\rmR}{{\rm R}}

\newcommand{\rmd}{{\rm d}}

\newcommand{\gam}{\gamma}
\newcommand{\Gam}{\Gamma}
\newcommand{\Dlt}{\Delta}

\newcommand{\GamL}{\Gamma_{\rm L}}
\newcommand{\GamR}{\Gamma_{\rm R}}
\newcommand{\dg}{\dagger}

\begin{document}
\draft

\title{Coulomb blockade double-dot Aharonov-Bohm interferometer:
      giant fluctuations}

\author{Feng Li}
\affiliation{State Key Laboratory for Superlattices and
Microstructures, Institute of Semiconductors,
Chinese Academy of Sciences, P.O.~Box 912, Beijing 100083, China}

\author{HuJun Jiao}
\affiliation{State Key Laboratory for Superlattices and
Microstructures, Institute of Semiconductors,
Chinese Academy of Sciences, P.O.~Box 912, Beijing 100083, China}

\author{JunYan Luo}
\affiliation{State Key Laboratory for Superlattices and
Microstructures, Institute of Semiconductors,
Chinese Academy of Sciences, P.O.~Box 912, Beijing 100083, China}

\author{Xin-Qi Li}
\affiliation{State Key Laboratory for Superlattices and
Microstructures, Institute of Semiconductors,
Chinese Academy of Sciences, P.O.~Box 912, Beijing 100083, China}
\affiliation{ Department of Physics, Beijing Normal University,
Beijing 100875, China }

\author{S.A. Gurvitz}
\affiliation{Department of Particle Physics,
 Weizmann Institute of Science, Rehovot 76100, Israel}

\date{\today}

\begin{abstract}
Electron transport through two parallel quantum dots is a kind of
solid-state realization of double-path interference. We demonstrate
that the inter-dot Coulomb correlation and quantum coherence would
result in strong current fluctuations with a divergent Fano
factor at zero frequency. We also provide physical
interpretation for this surprising result, which displays its
generic feature and allows us to recover this phenomenon in more
complicated systems.
\end{abstract}

\maketitle {\it Introduction}.---
As an analogue of Young's double-slit interference \cite{Fey70},
electron interfering through mesoscopic systems, e.g., a ring-like
Aharonov-Bohm (AB) interferometer with a quantum dot in one of the
interfering paths, is of interest for many fundamental reasons
\cite{Hac01463}. %%
The AB oscillation of conductance has been observed in both closed
\cite{Yac954047} and open geometry \cite{Buk98}, together with
elegant theoretical analysis \cite{Aha02}. Recently, further study
was carried out for the closed-loop setup, with particular focus on
the multiple-reflection induced inefficient ``which path"
information by a nearby charge detector \cite{Kang08}.

Going beyond the mere quantum interference, incorporation of Coulomb
correlation between the two paths should be of great interest. This
can be realized by transport through parallel double dots (DD) in
Coulomb blockade regime. For such DD setup, existing studies include
the cotunneling interference \cite{Los001035,Los-01,Kon013855,Sig06},
and two-loops (two fluxes) interference with the two dots as an
artificial molecule \cite{Hol0102,Jia02}. Remarkably, super-Poisson noise
and giant Fano Factor were predicted in this system, as generated by
the Coulomb correlations \cite{Los-01,Wan0703745,Dong08}.

It was very recently found \cite{Li0803} that the Coulomb
blockade in parallel dots pierced by magnetic flux $\Phi$ completely
blocks the resonant current for any value of $\Phi$ except for
integer multiples of the flux quantum $\Phi_0$. It was shown there
that this effects in a quantum analogue of self-trapping phenomenon
in non-linear systems. In the present paper we concentrate on Coulomb
blockade effects in parallel dots, where dephasing and lossy
channels are included. In particular we concentrate on the
shot-noise spectrum. We demonstrate that in the absence of dephasing
and lossy channels this quantity diverges at zero frequency. The
most important result of our analysis is an explanation of this
phenomenon using symmetry arguments. This explanation displays a new
way for a simple treatment of complicated Coulomb blockade effects
in the presence of quantum interference.

 \begin{figure}
 \center
 \includegraphics[scale=0.8]{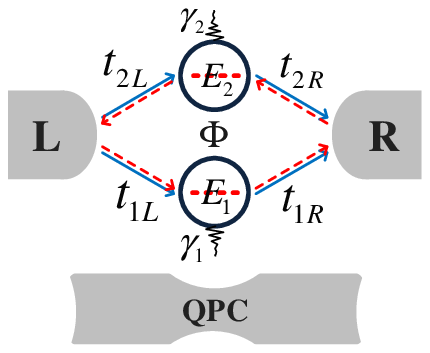}
 \caption{ (Color online)
 Schematic setup of a double-dot Aharonov-Bohm interferometer.
To address dephasing and electron-loss effects,
a nearby quantum-point-contact (QPC) detector and lossy channels
(with strength $\gamma_{1(2)}$) are introduced. }
 \end{figure}

%% ==================================================================
%Partly for this reason, to our knowledge,
%their comparative study is overlooked in literature.
%But, the electronic interferometer differs from
%the optical one in at least two aspects.
%%%
%First, in the optical double-slit interference, the photon arriving
%at one slit never comes back and is then scattered to another slit.
%In contrast, electron in the double-dot interferometer is subject
%to such kind of multiple forward-backward scattering.
%%%
%Second, despite Dirac's statement that
%a photon or an electron can interfere only with itself,
%a self-interfering electron, however,
%can interact with another self-interfering electron,
%while the self-interfering photons are independent to each other.
%%%
%In this work we demonstrate that these two additional complexities
%would result in novel features in the interference pattern,
%dephasing behavior, and fluctuation properties.
%
%
%
%\emph{Some points:}---
%%%
%\textbf{(1)} Dirac's statement:
%\emph{a photon or an electron can interfere only with itself}.
%This means that the interference pattern is determined by
%each individual electron.
%Then, how does the Coulomb interaction between electrons affect
%the interference pattern?
%It should be noted that the two interfering components \emph{cannot}
%have self-interaction.
%%%
%\textbf{(2)} ...
%
%%% ======================================================================

%%\section{Model Consideration}

{\it Model}.---
Consider double dots (DD) connected in parallel to two leads.
For simplicity we assume that in each of the dots
there is only one level, $E_{1}$ and $E_{2}$, involved in the transport.
Also, we neglect the spin degrees of freedom.
In case of strong Coulomb blockade, the effect of spin
can be easily restored by doubling the tunneling rates
of each QD with the left lead.
The system is described by the following Hamiltonian,
\begin{align}
H=H_0+H_T+\sum_{\mu =1,2} E_\mu d_\mu^\dagger d_\mu +Ud_1^\dagger
d_1 d_2^\dagger d_2\, . \label{a1}
\end{align}
Here the first term, $H_0=\sum_k [E_{kL}a_{kL}^\dagger a_{kL}
+E_{kR}a_{kR}^\dagger a_{kR}]$,
describes the leads and $H_T$ describes their coupling
to the dots,
\begin{align}
H_T=\sum_{\mu,k}\Big (t_{\mu L}d_\mu^\dagger a_{kL}
 +t_{\mu R}a_{kR}^\dagger d_\mu\Big )+{\rm H.c.}\, , \label{a2}
\end{align}
where $\mu=1,2$ and $a_{kL}^\dagger$ and~$a_{kR}^\dagger$ are the
creation operators for the electrons in the leads while
$d_{1,2}^\dagger$ are the creation operators for the DD. The last
term in Eq.~(\ref{a1}) describes the interdot repulsion. We assume
that there is no tunnel coupling between the dots and that the
couplings of the dots to the leads, $t_{\mu L(R)}$, are independent of energy.
In the absence of a magnetic field one can always choose
the gauge in such a way that all couplings are real.
In the presence of a magnetic flux $\Phi$, however, the tunneling
amplitudes between the dots and the leads are in general complex.
We write $t_{\mu L(R)}={\bar t}_{\mu L(R)}e^{i\phi_{\mu L(R)}}$,
where $\bar t_{\mu L(R)}$ is the coupling without the magnetic field.
The phases are constrained to satisfy
$\phi_{1L}+\phi_{1R}-\phi_{2L}-\phi_{2R}=\phi$, where $\phi\equiv
2\pi\Phi/\Phi_0$.

To account for dephasing effect, we introduce a ``which path" measurement
by a nearby point contact (PC) detector \cite{Buk98},
with a model description as in Ref.\ \onlinecite{Wan0703745}.
To make contact with conventional double-slit interferometer,
we also introduce electron lossy channels.
Slightly differing from Ref.\ \onlinecite{Aha02},
instead of the semi-infinite tight binding chain introduced there,
we model the lossy channels by attaching each dot with an
electronic side reservoir,
which is particularly suited in the master equation approach.
The side-reservoir model was originally proposed
by B\"uttiker in dealing with phase-breaking effect \cite{But8688},
i.e., electron would lose phase information
after entering the reservoir first, then returning back from it.
But here, we assume that the reservoir's Fermi level is much lower
than the dot energy.
As a result, electron only enters the reservoir
\emph{unidirectionally}, never coming back.

%\section{Formalism}
{\it Formalism}.---
The transport properties of the above described system,
both current and fluctuations,
can be conveniently studied by the number-resolved master equation
\cite{Gur96,Li05066803,Luo07085325}.
The central quantity of this approach is the number-conditioned state,
$\rho^{(n)}(t)$ of the double dots,
where $n$ is the electron number passed through the junction between
the DD and an assigned lead where number counting is done.
Very usefully, $\rho^{(n)}(t)$ is related to the electron-number
distribution function, in terms of $P(n,t)={\rm Tr}[\rho^{(n)}(t)]$,
where the trace is over the DD states. From $P(n,t)$ the current
and its fluctuations can be readily analyzed.
For current, it simply reads
$I(t)=ed \la n(t)\ra /dt$, where $\la n(t)\ra=\sum_n n P(n,t)$.
For current fluctuations, we employ the MacDonald's formula,
$S(\omega)=2\omega\int_{0}^{\infty} dt \sin\omega t \frac{d}{dt}
[\la n^2(t)\ra - (\bar{I}t)^2 ] $, to calculate the noise spectrum.
Here, $\la n^2(t)\ra=\sum_n n^2 P(n,t)$, and $\bar{I}$ is the
stationary current.
In practice, instead of directly solving $P(n,t)$, the reduced quantity
$\la n^2(t)\ra$ can be obtained more easily by constructing its
equation of motion, based on the ``$n$"-resolved master equation
\cite{Li05066803,Luo07085325}.

Under inter-dot Coulomb blockade, i.e., the DD can be
simultaneously occupied at most by one electron,
the Hilbert space of the DD state is reduced
to $|0\ra \equiv |00\ra$, $|1\ra \equiv |10\ra$, and $|2\ra \equiv |01\ra$,
where $|10\ra$ means the upper dot occupied and the lower dot unoccupied,
and other states have similar interpretations.
Following Ref.\ \onlinecite{Luo07085325}, the ``$n$"-resolved master equation
in this basis can be straightforwardly carried out as
%\begin{subequations}
%\bea
%\dot{\rho}_{00}&=&
%-4 \Gamma _L \rho _{00}
%+\left(\gamma +\Gamma _R\right) \rho _{11}
%+\left(\gamma +\Gamma _R\right) \rho _{22}   \nl
%&&+e^{-i \phi/2} \Gamma_R \rho_{12}
%+e^{i \phi/2} \Gamma _R \rho_{21}\\
%%%%%
%\dot{\rho}_{11}&=&
%2\Gamma _L \rho _{00}
%-\left(\gamma +\Gamma _R\right) \rho _{11}   \nl
%&&-\frac{1}{2} e^{-i \phi/2} \Gamma _R \rho_{12}
%-\frac{1}{2} e^{i \phi/2} \Gamma _R \rho_{21}\\
%%%%%
%\dot{\rho}_{22}&=&
%2\Gamma _L \rho _{00}
%-\left(\gamma +\Gamma _R\right) \rho _{22}   \nl
%&&-\frac{1}{2} e^{-i \phi/2} \Gamma _R \rho_{12}
%-\frac{1}{2} e^{i\phi/2} \Gamma _R \rho_{21}\\
%%%%%
%\dot{\rho}_{12}&=&
%e^{-i \phi/2} 2\Gamma _L \rho_{00}
%-\frac{1}{2} e^{i \phi/2} \Gamma _R (\rho _{11}+\rho_{22})   \nl
%&&-\frac{1}{2}  \left(\gamma_d+2 \gamma +2 i \Delta
%  +2 \Gamma _R\right) \rho_{12}
%\eea
%\end{subequations}

\begin{widetext}
\begin{subequations}
\bea
\dot{\rho}^{(n)}_{00}=
-2 \Gamma_L \rho_{00}^{(n)}
+\left(\gamma +\Gamma_R\right) \rho_{11}^{(n-1)}
+\left(\gamma +\Gamma_R\right) \rho_{22}^{(n-1)}
+e^{i \left(\phi_{\text{R1}}-\phi_{\text{R2}}\right)}
\Gamma_R \rho_{12}^{(n-1)}
+e^{i \left(\phi_{\text{R2}}-\phi_{\text{R1}}\right)}
\Gamma_R \rho_{21}^{(n-1)}
\eea
%2222
\bea
\dot{\rho}_{11}^{(n)}=
\Gamma_L \rho_{00}^{(n)}
-\left(\gamma +\Gamma_R\right) \rho_{11}^{(n)}
-\frac{1}{2} e^{i \left(\phi_{\text{R1}}-\phi_{\text{R2}}\right)}
\Gamma_R \rho_{12}^{(n)}
-\frac{1}{2} e^{i \left(\phi_{\text{R2}}-\phi_{\text{R1}}\right)}
\Gamma_R \rho_{21}^{(n)}
\eea
%3333
\bea
\dot{\rho}_{22}^{(n)}=
\Gamma_L \rho_{00}^{(n)}
-\left(\gamma +\Gamma_R\right) \rho_{22}^{(n)}
-\frac{1}{2} e^{i \left(\phi_{\text{R1}}-\phi_{\text{R2}}\right)}
\Gamma_R \rho_{12}^{(n)}
-\frac{1}{2} e^{i\left(\phi_{\text{R2}}-\phi_{\text{R1}}\right)}
\Gamma_R \rho_{21}^{(n)}
\eea
%44444
\bea
\dot{\rho}_{12}^{(n)}=
e^{i \left(\phi_{\text{L1}}-\phi_{\text{L2}}\right)}
\Gamma_L \rho_{00}^{(n)}
-\frac{1}{2} e^{i \left(\phi_{\text{R2}}-\phi_{\text{R1}}\right)}
\Gamma_R \rho_{11}^{(n)}
-\frac{1}{2} e^{i \left(\phi_{\text{R2}}-\phi_{\text{R1}}\right)}
\Gamma_R \rho_{22}^{(n)}
-\frac{1}{2}  \left(\gamma_d+2 \gamma +2 i \Delta +2 \Gamma_R\right)
\rho_{12}^{(n)}
\eea
%5555
\bea
\dot{\rho}_{21}^{(n)}=
e^{i \left(\phi_{\text{L2}}-\phi_{\text{L1}}\right)} \Gamma_L \rho_{00}^{(n)}
-\frac{1}{2} e^{i \left(\phi_{\text{R1}}-\phi_{\text{R2}}\right)}
\Gamma_R\rho_{11}^{(n)}
-\frac{1}{2} e^{i \left(\phi_{\text{R1}}-\phi_{\text{R2}}\right)}
\Gamma_R \rho_{22}^{(n)}
-\frac{1}{2}  \left(\gamma_d+2 \gamma -2 i \Delta +2 \Gamma_R\right)
\rho_{21}^{(n)}
\eea
\end{subequations}
\end{widetext}
In the above equations, $\Dlt=E_1-E_2$ is the offset of the dot
levels. $\Gamma_{L(R)}=2\pi D_{L(R)}|t_{L(R)}|^2$, and
$\gamma_{1(2)}=2\pi D_{1(2)}|t_{1(2)}|^2$, are the respective rates
for the couplings to the left and right leads, as well as to the
side reservoirs. $D_{L(R)}$ and $D_{1(2)}$ are the density of states
of the leads and reservoirs, while $t_{L(R)}$ and $t_{1(2)}$ are the
respective tunneling amplitudes. Note that in actual calculation
presented in this paper we replaced $\Gamma_L$ with $2\Gamma_L$
(c.f. \cite{Gur96,Luo07085325}). In this work we assume that
$\gamma_1=\gamma_2=\gamma$. Finally, $\gamma_d$ characterizes
dephasing between the two dots, resulting for instance from the
``which path" measurement by the point contact.

Note that the master equations~(3a)-(3e) include the off-diagonal
density-matrix elements, so that they explicitly display their
quantum mechanical nature. However these equations can be derived
from the many-body Scr\"odinger equation only in the infinite bias
limit in the presence of Coulomb blockade
\cite{Gur96,Li05066803,Luo07085325}.

%% --------------------------------------------------------------------

\begin{figure}
 \center
 \includegraphics[scale=0.6]{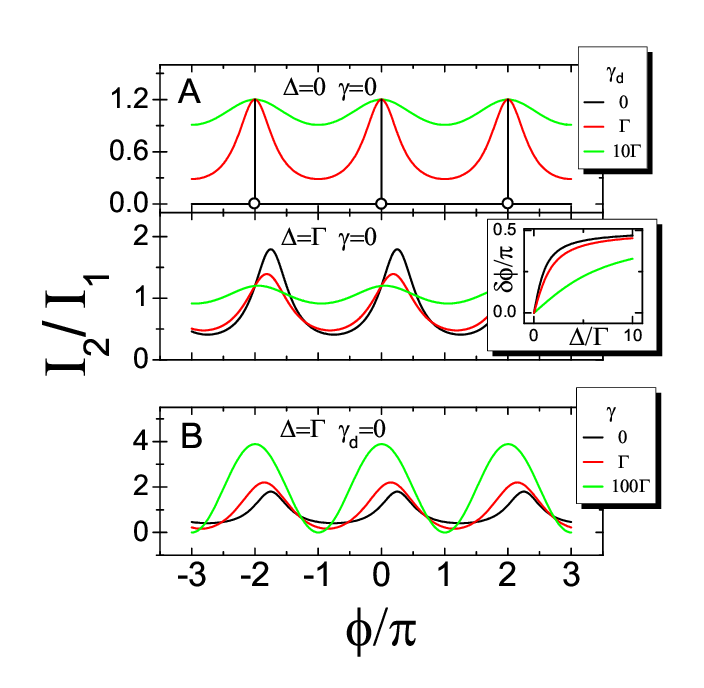}
 \caption{(Color online)
(A) Current switch and dephasing effect under closed geometry
($\gamma=0$) for aligned DD levels ($\Delta=0$).
(B) Phase shift for $\Delta\neq 0$ and electron loss effect.
With increasing the lossy strength $\gamma$,
the conventional double-slit interference pattern is recovered.}
 \end{figure}

%\section{Current: Interference Pattern}

{\it Current}.---
First, we consider the case without electron loss, i.e., $\gamma=0$.
Simple expression for the steady-state current is extractable:
 \be\label{I-2d-1}
 I=\left\{\frac{2(\gam_\rmd\!+\!2\Gam_\rmR)
 (1\!-\!\cos\phi)
 -4\Delta\sin\phi}{\gam_\rmd(\gam_\rmd+2\Gam_\rmR)
 +4\Delta^2}+\frac{1}{I_0}\right\}^{-1},
 \ee
where $I_0=4\Gam_\rmL\Gam_\rmR /(4\Gam_\rmL+\Gam_\rmR)$,
is the current in the absence of magnetic flux.
However, in the following we will use the current of
transport through a Coulomb-blockade single dot,
$I_1=2\GamL\GamR/(2\GamL+\GamR)$, to scale
the double-dot current, in order to highlight the interference features.

%\subsection{Current Switching}

For the limiting case $\gamma_d$=0, i.e., no dephasing between the two dots,
from \Eq{I-2d-1} we have
$I= I_0 \Delta^2/\{\Delta^2+I_0[\Gamma_R(1-\cos\phi)-\Delta \sin\phi]\}$.
Then a novel switching effect follows this result:
as $\Delta\rightarrow 0$, $I=I_0$ for $\phi=2\pi n$,
while $I=0$ for any deviation of $\phi$ from these values.
This remarkable behavior can be explained
by defining new basis states of the DD,
$d^{\dagger}_\mu|0\rangle\to\tilde d^{\dagger}_\mu |0\rangle$,
chosen such that
$\tilde d^\dagger_2 |0\rangle$ is not coupled to the right
reservoir, {\em i.e.}, $t_{2R}\to \tilde t_{2R}=0$,
then the current
would flow only through the state $\tilde d_{1}^\dagger|0\rangle$.
This can be realized by the unitary transformation \cite{Li0803}
\begin{align}
\left (\begin{array}{c}\widetilde {d}_{1}\\
\widetilde {d}_{2}\end{array}\right)={1\over {\cal
N}}\left(\begin{array}{cc}t_{1R}
& t_{2R}\\
-t^*_{2R}&t^*_{1R}\end{array}\right)\left (\begin{array}{c}d_{1}\\
d_{2}\end{array}\right), \label{a11}
\end{align}
with ${\cal N}=(\bar t_{1R}^2+\bar t_{2R}^2)^{1/2}$, which indeed
results in $\widetilde t_{2R}=0$.
Also, the coupling of $\widetilde d^\dagger_2|0\rangle$
to the left lead reads
\begin{align}
\widetilde t_{2L}(\phi)=-e^{i(\phi_{2L}-\phi_{1R})}(\bar t_{1L}\bar
t_{2R}\, e^{i\phi}-\bar t_{2L}\bar t_{1R})/{\cal N} .\label{a12}
\end{align}
It follows from this expression that $\widetilde t_{2L}=0$ for
$\phi=2n\pi$ provided that
$\bar t_{1L}/\bar t_{2L}=\bar t_{1R}/\bar t_{2R}$, or for
$\phi=(2n+1)\pi$ if $\bar t_{1L}/\bar t_{2L}=-\bar t_{1R}/\bar t_{2R}$.
Obviously, for noninteracting DD, $\widetilde d^\dagger_2|0\rangle$
has no contribution to current, while $\widetilde d^\dagger_1|0\rangle$
carries a magnetic-flux modulated current.
In the case of inter-dot Coulomb blockade,
however, whether the coupling of $\widetilde d^\dagger_2|0\rangle$
to the left lead is zero becomes of crucial importance.
If $\widetilde t_{2L}\neq 0$, then the state
$\widetilde d_1^\dagger |0\rangle$, carrying the current, will be
blocked by the inter-dot Coulomb repulsion.
As a result, the total current {\em vanishes}.
However, if the state $\widetilde d_2^\dagger |0\rangle$
is decoupled from {\em both\/} leads, it
remains unoccupied, so that the current can flow through the state
$\widetilde d_1^\dagger |0\rangle$. As shown above, this takes place
precisely for $\bar t_{1L}/\bar t_{2L}=\pm\,\bar t_{1R}/\bar
t_{2R}$. If this condition is not fulfilled, the current is always
zero, even for $\phi =2\pi n$.

As we demonstrated above, the switching effect
becomes very transparent in the particular basis of the DD states.
Still, it is very surprising how such a basis emerges
dynamically? Indeed, an electron from the left lead can
enter the DD system in any of SU(2) equivalent
superpositions of its states.
Therefore there exists a probability
for each electron to enter the DD in the superposition
that eliminates one of the links with the right lead. When
it happens, the electron would be trapped in this state.
Even if the probability of this event for one electron is
very small, the total number of electrons passing through
the DD goes to infinity for $t\rightarrow\infty$.
Therefore the trapping event
is always realized for large enough time.
In the presence of Coulomb blockade this would lead to
the switching effect, as explained above.

In the presence of dephasing, which is modelled by a which-path
detection in this work, the switching effect discussed above
will be smeared out, as shown in Fig.\ 2(A).
In appearance, the resultant interference pattern resembles
the usual one of the double-slit interferometer.
However, both qualitatively and quantitatively, there exists
remarkable differences, e.g., the unchanged current
at $\phi=2\pi n$, which is also the fully dephased current.

As the DD levels deviate from alignment, i.e., $\Delta\neq 0$,
the current switching phenomena will also disappear, as shown in Fig.\ 2(B).
Similar to dephasing,
from Eq.\ (4), we see that the current at $\phi=2\pi n$ is unaffected
by $\Delta$, too. However, for $\Delta\neq 0$, this is not the maximal current.
Accordingly, a phase shift of the interference pattern is implied.
In more generic sense,
this is nothing but the breaking of phase locking \cite{But86},
for two-terminal transport which can appear {\it only}
under finite bias voltage and typically
in the presence of electron-electron interactions \cite{Kon013855}.

In Fig.\ 2(B) we display also the effect of electron loss.
That is, we introduce lossy channels to make the interferometer
more and more open.
As a result, we see that all the above distinguished features
disappear and the conventional double-slit interference pattern
is restored by increasing the lossy strength $\gamma$.
The basic reason is that, as the dots become increasingly open,
the side reservoirs would reduce the occupation probability on the dots,
thus make the Coulomb correlation and back-reflection less important.

%\section{Current Fluctuations}
{\it Current Fluctuations}.---
Current fluctuations are usually characterized by the
zero-frequency shot noise, which can be calculated
by the particle-number-resolved master equation approach,
using the MacDonald's formula as sketched in the formalism.
Strikingly, for the present Coulomb blockade DD interferometer,
we find that
the zero-frequency shot noise can be highly super-Poissonian,
and can even become divergent as $\Delta\rightarrow 0$.
In the following we first demonstrate this novel result,
then show more other features of the noise.

%\subsection{Divergent Noise}

For coherent DD interferometer, analytical result of the frequency-dependent
noise can be obtained using the MacDonald's formula:
\begin{widetext}
\begin{align}\label{S-R}
S(\omega)= \frac{8\Gamma_L\Gamma_R [ 2\Gamma_L\Gamma_R\Delta^2
    -\Delta^4+3\Delta^2\omega^2 -2\omega^2(\Gamma^2_R +\omega^2)]\bar{I}}
    {[(2\Gamma_L+\Gamma_R)\Delta^2 - (2\Gamma_L+3\Gamma_R)\omega^2]^2
      +\omega^2(2\Gamma_L\Gamma_R+2\Gamma^2_R+\Delta^2-\omega^2)^2}
    + 2\bar{I} ~.
\end{align}
\end{widetext}
Here we have assumed $\phi=2\pi n$.
At zero frequency limit, the Fano factor reads
\begin{align}\label{F-1}
F\equiv\frac{S(0)}{2\bar{I}}
= \frac{  8\Gamma^2_L\Gamma^2_R + (4\Gamma^2_L+\Gamma^2_R) \Delta^2  }
{ (2\Gamma_L+\Gamma_R)^2 \Delta^2 }.
\end{align}
Strikingly, as $\Delta\rightarrow 0$, it becomes divergent!
Note that this divergence is not caused by the average current
$\bar{I}$, but the zero-frequency noise itself.
Very interestingly, from \Eq{S-R},
we find that the limiting order of
$\Delta\rightarrow 0$ and $\omega\rightarrow 0$,
would lead to different results.
That is, if we first make $\Delta\rightarrow 0$,
then $\omega\rightarrow 0$, the result reads
\begin{align}\label{F-2}
F=\frac{\Gamma^2_L+\Gamma^2_R}{(\Gamma_L+\Gamma_R)^2},
\end{align}
which is finite and coincides with the Fano factor of single-level
transport \cite{Luo07085325}.
The limiting order leading to \Eq{F-2} implies that we are considering
the noise for aligned DD levels.
In this case, as constructed above, see \Eq{a11} and Fig.\ 3(a),
the two transformed dot-states are decoupled to each other,
and one of them also decoupled to both leads if $\phi=2\pi n$.
As a result, equivalently, the transport is through a single channel,
leading to the Fano factor \Eq{F-2}.

\begin{figure}
\center
 \includegraphics[scale=0.7]{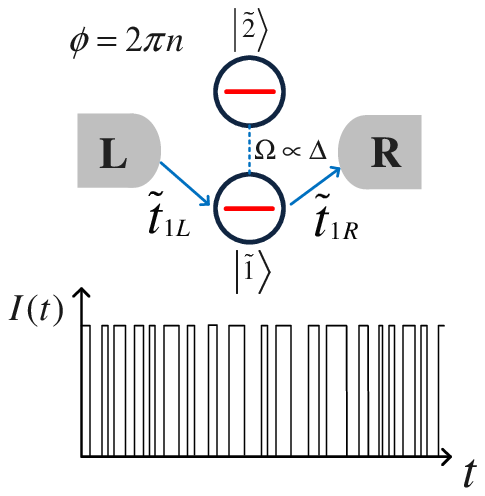}
 \caption{ (Color online)
 Schematic interpretation for the noise divergence.
\emph{Upper panel}: Effective coupling of the DD to the leads and
between the dots, in the representation of transformed DD states, i.e.,
$|\ti{1}\ra\equiv\ti{d}^{\dg}_1  |0\ra$ and
$|\ti{2}\ra\equiv\ti{d}^{\dg}_2  |0\ra$.
\emph{Lower panel}: Coarse-grained temporal current, with a telegraphic
noise nature which causes divergence of the zero-frequency noise
when $\Delta\rightarrow 0$ . }
 \end{figure}

However, for $\Delta \rightarrow 0$ but $\neq 0$, the situation is
subtly different. In this case, the two transformed states are
weakly coupled, with a strength $\propto\Delta$. Thus, the
transporting electron on state $\tilde{d}_1^{\dg}|0\ra$ can
occasionally tunnel to $\tilde{d}_2^{\dg}|0\ra$, which is
disconnected to both leads, and its occupation will block the
current until the electron tunnels back to $\tilde{d}_1^{\dg}|0\ra$
and arrives at the right lead. %%
Typically, this strong {\it bunching behavior}, induced by the interplay
of Coulomb interaction and quantum interference,
is well characterized by a profound super-Poissonian statistics.
In Fig.\ 3(b), the coarse-grained temporal current
with a telegraphic noise nature is plotted schematically.
We see that, as $\Delta \rightarrow 0$, the current switching would become
extremely slow, leading to very long time ($\sim 1/\Delta$) correlation
between the transport electrons.
It is right this long-time-scale fluctuation, or equivalently, the
low frequency component filtered out from the current, which causes
divergence of the shot noise as $\Delta \rightarrow 0$. This is
similar, in certain sense, to the well known $1/f$ noise, which goes
to divergence as $f \rightarrow 0$.

It is quite interesting that a similar divergence of the
noise-spectrum at zero frequency has been found for rather broad
conditions (but only for inelastic cotunneling regime) in the
framework of classical master equations. The latter neglects the
off-diagonal elements of the density matrix and assumes weak enough
tunneling \cite{Los-01}. In contrast, our quantum rate equations
approach goes beyond these assumptions. On the other hand it shows
that the switching effect and divergency of the noise-spectrum can
take place only at the large bias voltage \cite{Li0803}. Indeed, by
applying the unitary transformation (\ref{a11}), one can always
decouple one of the states from the right reservoir. However, one
still needs the total occupation of this state at $t\to\infty$.
Otherwise the Coulomb blockade is not complete and so the switching
effect. This condition can be realized only in the large bias limit,
where the energy levels of the dots are far from the corresponding
Fermi energy.

%%\subsection{Magnetic-Flux Dependence}
{\it Magnetic-Flux Dependence}.---
The previous study was restricted to zero magnetic flux so that
only the state $\tilde{d}_{1}^{\dg}|0\ra$ is connected to the leads.
Now we proceed to nonzero magnetic flux and $\Delta\neq 0$.
In this case, $\tilde{\Gamma}_{1L}(\phi)$ and $\tilde{\Gamma}_{2L}(\phi)$
are nonzero in general.
By tuning the flux from $\phi=0$ to $\pi$,
the effective coupling $\tilde{\Gamma}_{2L}$ is switched on,
while $\tilde{\Gamma}_{1L}$ switched off.
As a result, the strong current fluctuation at $\phi=0$
is considerably suppressed owing to this transition
to transport through $\tilde{d}_{1}^{\dg}|0\ra$ and
$\tilde{d}_{2}^{\dg}|0\ra$ in series.
In between $\phi=0$ and $\pi$, however, we find a local minimum
for the Fano factor, with a common value given by \Eq{F-2},
being independent of $\Delta$.
Its location in $\phi$, however, depends on $\Delta$.
This is because,
for different $\Delta$, one can always find a proper $\phi$,
such that $\tilde{d}_{1}^{\dg}|0\ra$ couples to the left lead,
both directly and indirectly through $\tilde{d}_{2}^{\dg}|0\ra$,
with an effective coupling strength $\Gamma_{L}$.
Remind also that, the coupling of $\tilde{d}_{1}^{\dg}|0\ra$
to the right lead is $\sim \Gamma_R$.
Accordingly, the Fano factor of \Eq{F-2} is reached.

\begin{figure}
\center
 \includegraphics[scale=0.7]{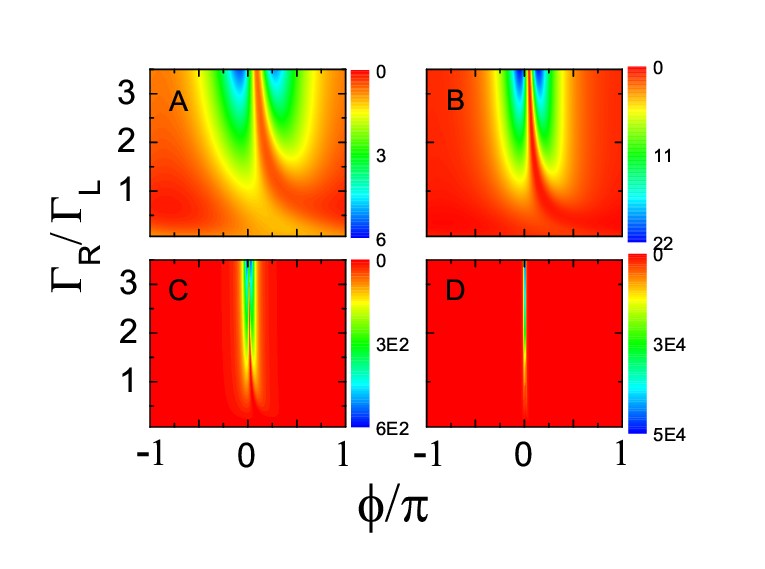}
 \caption{ (Color online) Contour plot of the Fano factor
\emph{versus} the (scaled) magnetic flux $\phi$
and tunnel-coupling asymmetry $\Gamma_R/\Gamma_L$,
for different DD level detuning:
$\Delta=\Gamma$ (A), $0.5\Gamma$ (B),  $0.1\Gamma$ (C),
and $0.01\Gamma$ (D). }
 \end{figure}

In Fig.\ 4, we display the Fano factor versus the (scaled)
magnetic flux (phase difference $\phi$)
and the tunnel-coupling asymmetry $\Gamma_R/\Gamma_L$.
Besides the $\phi$-dependence,
we see that, with the increase of $\Gamma_R/\Gamma_L$,
the Fano factor is considerably enhanced and becomes
highly super-Poissonian.
Interpretation for this dependence is referred to
Ref.\ \onlinecite{Wan0703745}, where the concept of
effective fast-to-slow channels was proposed.

Finally, not shown in Fig.\ 4 includes the effects of dephasing
and electron loss.
It is clear that, for dephased (original) dots,
we can no longer construct
the superposition states $\tilde{d}_{1}^{\dg}|0\ra$ and
$\tilde{d}_{2}^{\dg}|0\ra$.
Then, the fast-to-slow channel induced bunching behavior
is not anticipated,
and the Fano factor is reduced to the Poissonian value.
For lossy effect, we conclude that,
with increasing the lossy strength ($\gamma$),
shorter duration time on dots will weaken the role of Coulomb interaction
and multiple reflections, making the noise characteristics
Poissonian, like that from usual random emission.

\vspace{1cm} {\it Note Added.}--- After the submission of present
work to the arXiv:0812.0846-eprint, a very recent paper by Urban and
K\"onig was caused into our attention \cite{Urb08}, where the
enhancement of shot noise and even divergence were found in the
absence of inter-dot but in the presence of intra-dot Coulomb
blockade. In that case, the electron spin plays an essential role.
In our DD Coulomb blockade regime, however, the electron spin is
irrelevant to the super-Poisson noise and its divergence.

%\vspace{2cm}
%\begin{acknowledgments}
{\it Acknowledgments.}---
This work was supported by the National Natural Science Foundation
of China under grants No. 60425412 and No. 90503013.
X.Q.L. acknowledges the
Albert Einstein Minerva Center for Theoretical Physics
for partially supporting his visit to the Weizmann Institute of Science.
S.G is grateful to the Max Planck Institute for the Physics of Complex
Systems, Dresden, Germany for kind hospitality.
%\end{acknowledgments}

%\clearpage

%\bibliography{D:/bibliography/bibrefs}

\end{document}